\documentclass[letterpaper,twocolumn,10pt]{article}
\usepackage{usenix,epsfig,endnotes}
\usepackage{tugraz_defaults}
\usepackage{csquotes}

\usepackage{tugraz_defaults}
\usepgfplotslibrary{dateplot}
\usepgfplotslibrary{fillbetween}

\newcommand{\parhead}[1]{\noindent\textbf{#1.}\ }

\begin{document}

\date{}

\title{Page Cache Attacks}

\author{{\rm Daniel Gruss}$^1$, {\rm Erik Kraft}$^1$, {\rm Trishita Tiwari}$^2$, {\rm Michael Schwarz}$^1$,\\{\rm Ari Trachtenberg}$^2$, {\rm Jason Hennessey}$^3$, {\rm Alex Ionescu}$^4$, {\rm Anders Fogh}$^5$ \\
$^1$ Graz University of Technology, $^2$ Boston University, $^3$ NetApp, $^4$ CrowdStrike, $^5$ Intel Corporation}

\maketitle

\pagestyle{empty}

\subsection*{Abstract}
We present a new hardware-agnostic side-channel attack that targets one of the most fundamental software caches in modern computer systems:
the operating system page cache. 
The page cache is a pure software cache that contains all disk-backed pages, including program binaries, shared libraries, and other files, and our attacks thus work across cores and CPUs.
Our side-channel permits unprivileged monitoring of some memory accesses of other processes, with a spatial resolution of \SI{4}{\kilo\byte} and a temporal resolution of \SI{2}{\micro\second} on Linux (restricted to $6.7$ measurements per second) and \SI{466}{\nano\second} on Windows (restricted to $223$ measurements per second); this is roughly the same order of magnitude as the current state-of-the-art cache attacks.
We systematically analyze our side channel by demonstrating different local attacks, including a sandbox bypassing high-speed covert channel, timed user-interface redressing attacks, and an attack recovering automatically generated temporary passwords.
We further show that we can trade off the side channel's hardware agnostic property for remote exploitability. 
We demonstrate this via a low profile remote covert channel that uses this page-cache side-channel to exfiltrate information from a malicious sender process through innocuous server requests. 
Finally, we propose mitigations for some of our attacks, which have been acknowledged by operating system vendors and
slated for future security patches. 

\section{Introduction}\label{sec:intro} 
Modern processors are highly optimized for performance and efficiency.
A large share of these optimizations is based upon \emph{caching} - taking advantage of temporal and spatial locality to minimize slower memory or disk accesses. 
Indeed, caching architectures typically fetch or prefetch code and data into fast buffers closer to the processor. 

Although side-channels have been known and utilized primarily in military contexts for decades~\cite{Young2002,Lampson1973}, the idea of cache side-channel attacks gained more attention over the last twenty years~\cite{Kocher1996,Bernstein2004,Percival2005}.
Osvik~\etal\cite{Osvik2006} showed that an attacker can observe the cache state at the granularity of a cache set using \PrimeProbe, and later Yarom~\etal\cite{Yarom2014} showed this with cache line granularity using \FlushReload. 
While different cache attacks have different use cases, the accuracy of \FlushReload remains unrivaled.

Indeed, virtually all \FlushReload attacks target pages in the so-called page cache~\cite{Yarom2014,Irazoqui2015Lucky,Gruss2015Template,Inci2016,Lipp2016,Irazoqui2016Cross}.
The page cache is a pure software cache implemented in all major operating systems today, and it contains virtually all pages in use. 
Pages that contain data accessible to multiple programs, such as disk-backed pages (\eg program binaries, shared libraries, other files, etc.), are shared among all processes regardless of privilege and permission boundaries~\cite{Gorman2004}. The operating system uses the page cache to store frequently used pages in memory, this obviating slow disk loads whenever a process needs to access said pages. 
There is a large body of works exploiting \FlushReload in various scenarios over the past several years~\cite{Yarom2014,Irazoqui2015Lucky,Gruss2015Template,Inci2016,Lipp2016,Irazoqui2016Cross}. 
There have also been a series of software (side-channel) cache attacks in the literature, including attacks on the browser cache~\cite{Felten2000,Jackson2006,Bortz2007,Jia2015,VanGoethem2015} and exploiting page deduplication~\cite{Suzaki2011,Owens2011,Xiao2012covert,Xiao2013security,Barresi2015,Gruss2015dedup,Bosman2016,Razavi2016}; however, page deduplication is mostly disabled or limited to deduplication within a security domain today~\cite{WindowsServerDeduplication, RedHatKSM, Vanderveen2016}.

In this paper, we present a new attack on the operating system page cache. 
We present a set of local attacks that work entirely without any timers, utilizing operating system calls (\texttt{mincore} on Linux and \texttt{QueryWorkingSetEx} on Windows) to elicit page cache information. 
We also show that page cache metadata can leak to a remote attacker over a network channel, producing a stealthy covert channel between a malicious local sender process and an external attacker.

We comprehensively evaluate and characterize the software-cache side channel by comparing it to hardware-cache side channels.
Like the recent DRAMA attack~\cite{Pessl2016,Wang2017leaky}, our side-channel attack works across cores and across CPUs
with a spatial granularity of \SI{4}{\kilo\byte}. 
For comparison, the spatial granularity of the DRAMA attack is \SI{2}{\kilo\byte} on dual-channel systems up to and including the Haswell processor architecture, and \SI{1}{\kilo\byte} on more recent dual-channel systems.
The temporal granularity of the DRAMA attack is around \SI{300}{\nano\second}, whereas the temporal granularity of our attack is \SI{2}{\micro\second} on Linux (restricted to $6.7$ measurements per second) and \SI{466}{\nano\second} on Windows (restricted to $223$ measurements per second). 
Hence, we conclude that our attack can compete with the current state-of-the-art in microarchitectural attacks.

Finally, we present several ways to mitigate our attack in software, and observe that certain page replacement algorithms reduce the applicability
of our attack while simultaneously improving the system performance. 
In our responsible disclosure, both Microsoft and the Linux security team acknowledged the problem and informed us that they will follow our recommendations with security patches to mitigate our attack.

To summarize, we make the following contributions:
\begin{compactenum}
	\item We present a novel attack targeting the page cache.
	\item We present a high-speed covert channel which is agnostic to specific hardware configurations.
	\item We present a set of local attacks which can compete with state-of-the-art microarchitectural attacks.
	\item We present a remote attack which can leak information across the network.
\end{compactenum}

We begin in \Cref{sec:background} with background information on hardware caches, cache attacks, and software caches,
followed by our threat model in \Cref{sec:threatmodel}.
\Cref{sec:attack} overviews our attack.
\Cref{sec:step1} presents a novel method to spy on the page cache state.
\Cref{sec:eviction} shows how page cache eviction can be done efficiently on Linux and Windows.
\Cref{sec:local} presents (timing-free) local page cache attacks.
\Cref{sec:remote} presents remote page cache attacks.
\Cref{sec:countermeasures} discusses different countermeasures against our attack.
\Cref{sec:conclusion} concludes our work.

\section{Background}\label{sec:background} 
We begin with a brief discussion of hardware and software cache attacks, followed
by some background on the operating system page cache that we exploit.

\subsection{Hardware and Software Cache Side-Channel Attacks}\label{sec:bg_cat}
The suggestion of cache attacks harks back to the timing attacks of Kocher~\cite{Kocher1996}.
Osvik~\etal\cite{Osvik2006} presented a technique with a finer granularity called \PrimeProbe.
Yarom~\etal\cite{Yarom2014} presented \FlushReload, which is still today the cache attack technique with the highest accuracy (virtually no false negatives or false positives) and a finer granularity than most other attacks (one cache line).
Consequently, \FlushReload is also used in other applications, including the covert channel in Spectre~\cite{Kocher2019} and Meltdown~\cite{Lipp2018meltdown}. 
\FlushReload requires shared memory with the victim application.
However, all modern operating systems share code and unmodified data of every program and shared library (and any unmodified file-backed page in general) across privilege boundaries and applications.

Caches also exist in software, caching remote data, data that has been retrieved from slow or offline storage, or precomputed results.
Some of these caches have very specific use-cases, such as browser caches used for website content; other caches are more generic,
such as the page cache that stores a large portion of code and data used. 
Caches make use of the principle of locality to retain common computations closer to the processor, and consequently
they can leak information about the cache contents.

For example, browser caches leak information about browsing history and other possibly sensitive user information~\cite{Felten2000,Jackson2006,Bortz2007,Jia2015,VanGoethem2015}.
Requested resources may have different access times, depending on whether the resource is being served from
a local cache or a remote server, and these differences can be distinguished by an attacker.
As another example of a software-based side channel, page-deduplication attacks exploit page deduplication across security boundaries.
A copy-on-write page fault reveals the fact that the requested page was deduplicated and that
another process must have a page with identical content.  
Suzaki~\etal\cite{Suzaki2011} presented the first page-deduplication attack, which detected programs running in co-located virtual machines.
Subsequently, several other page-deduplication attacks were demonstrated~\cite{Owens2011,Xiao2012covert,Xiao2013security,Gruss2015dedup}.
Today, page deduplication is either completely disabled for security reasons or restricted to deduplication within a security domain~\cite{WindowsServerDeduplication, RedHatKSM, Bosman2016,Razavi2016}.

\subsection{Operating System Page Cache}\label{sec:pagecache}

Virtual memory creates the illusion for each involved process of running alone on the system.  To do this,
it provides isolation between processes so that different processes may operate on the same addresses
without interfering with each other.  Each virtual memory page may be mapped by the operating
system, with varying properties, to an arbitrary physical memory page.

When multiple processes map a virtual page to the same physical page, this page is part of
\emph{shared memory}.  Shared memory typically may arise out of inter-process communication or, more broadly, to reduce physical memory consumption.  For example, if shared library and common binary pages
on the hard disk are mapped multiple times by different processes, they map to the same pages in physical
memory.

Indeed, any page that might be used by more than one process may be mapped as shared memory.  However,
if a process wants to write to such a page, it must first secure a private copy of the page, so as not to
break the isolation between processes.  The efficiency savings come because a great many pages are never
modified and, instead, remain shared among multiple processes in a read-only state.

The operating system page cache is a generalization of the above memory sharing scenario, and, in fact, all modern operating systems (\eg Windows, Linux, and OS X) implement a page cache.  The page cache contains all pages that are memory mapped files, any file read from
the disk, and (depending on the system) possibly other pages such as anonymous pages or shared memory~\cite{Gorman2004}.
The operating system keeps track of which pages in the page cache are clean (\ie their data is unmodified from the disk version)
and which are dirty (\ie modified since they were first loaded from the disk).  
Ideally, the page cache incorporates all available memory, allowing the operating system to minimize the disk I/O.

The introduction of a page cache disrupts the traditional functioning of the operating system under a page fault.
Without a page cache, the operating system reserves a free physical page frame, loads the data from the disk into that
physical page frame, and then maps a virtual page to the physical page frame accordingly.
If there are no available physical page frames, the system swaps out pages to the disk using an operating system-dependent
page-replacement algorithm. 
In Linux, this algorithm had traditionally been based on a variant of the Least Recently Used (LRU) paradigm~\cite{Corbato1968paging}, and
LRU-related data structures can still be found throughout the kernel code. More recent Linux versions implement an improved variant called CLOCK-Pro~\cite{Jiang2005} along with several adaptions~\cite{LWN_clockpro}.  Within this improved framework,
Linux moves pages among multiple lists (an inactive list, an active list, and a recently evicted list).
In contrast to Linux, Windows uses the \emph{working-set} model of page caching to introduce more fairness among processes
competing for memory~\cite{Denning1968,Denning1980,Carr1981}.
The page replacement algorithm used on Windows was based on Clock or pseudo-random replacement~\cite{Friedman1999,Russinovich1998} in older Windows versions, and today is likely a variant of the Aging algorithm~\cite{Bruno2013technet}. 

With a page cache, the operating system endeavors to make full use of all physical page frames, and a page-replacement
algorithm is still needed for evicting page cache pages (swapping is less relevant on modern operating systems~\cite{DebianSSD,Horn2017Swap,Crowthers2011}). 
Also pages from KVM virtual machines are cached in the host-side page cache if the machine is configured to use a write-back caching strategy~\cite{Djordjevic2015}.

Both Linux and Windows provide mechanisms for checking whether a page is
resident in the page cache - the \texttt{mincore} system call for Linux, and the \texttt{QueryWorkingSetEx} system call for Windows.

\section{Threat Model}\label{sec:threatmodel}
Our threat model is based on the threat model for \FlushReload~\cite{Yarom2014,Irazoqui2015Lucky,Gruss2015Template,Inci2016,Lipp2016,Irazoqui2016Cross}.

Specifically, we assume that attacker and victim have access to the same operating system page cache.
On Linux, we also assume that the attacker has read access to the target page, which may be any page of any attacker-accessible file on the system.
This assumption is satisfied, for example, when attacker and victim are
\begin{compactitem}
\item processes running under the same operating system, or
\item processes running in isolated sandboxes with shared files (\eg Firejail~\cite{Firejail}). 
\end{compactitem}
On Windows, read access to the target page is not necessary for our attack.

Our local attacks are timing-free, in that they do not rely on hardware timing differences.
Our remote attack leverages timing differences between memory and disk access, measured on a remote system, as a proxy for the required local information.

\section{High-Level View of the Attack}\label{sec:attack}
Our attack fundamentally relies on the attacker's capability to distinguish whether a page is in the page cache or not.
In the local attack we are agnostic to the underlying hardware, \ie we do not exploit any timing differences although this would be practically possible on virtually all systems.
Thus, we use the \texttt{mincore} system call on Linux for this purpose and the \texttt{QueryWorkingSetEx} system call on Windows. 
The \texttt{mincore} system call returns which pages of a memory range are present in memory (\ie in the page cache) and which are not. 
Likewise, the \texttt{QueryWorkingSetEx} system call returns a list of pages that are in the current working set of a process, and thus are present in the page cache. 

Bringing the page cache into a known state is not trivial, as it behaves like a fully associative cache.
Previous approaches for page cache eviction can lead to out-of-memory situations~\cite{Seaborn2015BH,Gruss2016Row,Vanderveen2016} or consume too much time and impose system pressure~\cite{Gruss2018Rowhammer}.
This is not practical when evicting pages often, \eg multiple times per second.
Hence, they have not been used in published side-channel attacks so far, but only to support other attacks, \eg relocation of a page for Rowhammer.
For Linux, we devise a working-set-based eviction strategy that efficiently accesses groups of other pages more frequently than the page to evict. 

On Windows, our attack is much more efficient than on Linux.
On Linux, the page cache is directly influenced by all processes.
In contrast, Windows has per-process working sets~\cite{WinAPI}, and the page cache is influenced indirectly through these working sets. Hence, for Windows, we present an attack which evicts pages only from the working set of the victim process, but not from the page cache (\ie not from DRAM), \ie causing no additional disk accesses.
Although both attack variants follow the same attack methodology, we have to distinguish between the Linux and Windows variant at several places in the remainder of the paper. 

In contrast to hardware cache attacks and page-deduplication attacks, our local attacks are non-destructive, allowing us to repeat measurements.
Measuring whether a memory location is cached or not manipulates the state such that the information is not available anymore at a later point in both hardware cache attacks~\cite{Osvik2006,Yarom2014} and page-deduplication attacks~\cite{Suzaki2011,Owens2011}.
However, it is not the case for our local attack.
As we rely on the \texttt{mincore} and \texttt{QueryWorkingSetEx} system calls, we can arbitrarily check whether the page is in the page cache (on Linux) or the process working set~\cite{WinAPI} (Windows).
These checks are non-destructive as they neither modify nor influence the state of the page cache or the process working set with respect to the target memory location.

\begin{figureA}[t]{strategy_overview}[Attack overview.]
 \includegraphics[width=\hsize]{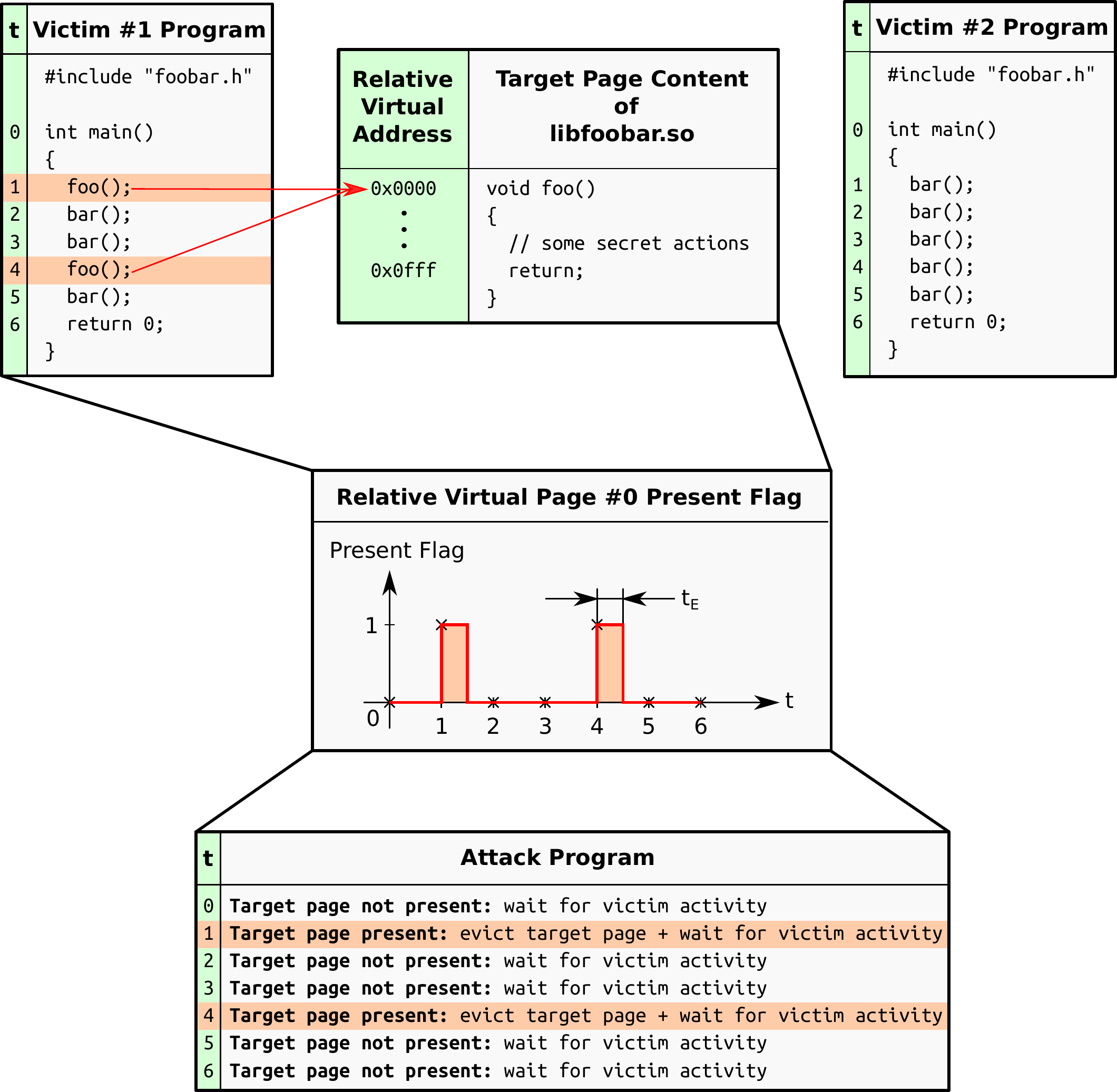}
\end{figureA}

Our attack is illustrated in \cref{fig:strategy_overview}.
The attacker wants to measure when the function \texttt{foo()} is called by a victim program.
The attacker determines the page which contains the function \texttt{foo()}.
By observing when the page is in the page cache, the attacker learns when \texttt{foo()} was called.

Our attack continuously runs through the following steps:
Initially, the target pages are in the page cache (on Linux) respectively the working set of the victim process (on Windows).
After the eviction, the page is not in the page cache (Linux) or process working set (Windows) anymore.
The attacker can now continuously probe when the page is added back in.
As soon as the page is found in the page cache (Linux) or the process working set (Windows), the attacker logs the memory access and evicts the page again.

In the following sections, we detail the two main steps of the attack, \ie determining the page cache state (defining the temporal resolution) and performing the page cache eviction (defining the maximum frequency at which the attack can be performed).

\section{Determining the Page Cache State}\label{sec:step1}
In this section, we discuss how to determine the page cache state. 
Note that although our attack starts with the page cache eviction, following the attack description is easier when understanding how to determine the page cache state first.

The attacker wants to determine when a specific page from a shared library is loaded into the page cache, as this is exactly the time of the access by the victim program.
Thus, the shared library containing the target page an attacker wants to observe accesses to has to be mapped into the attacker's address space.
This is possible using \texttt{mmap} on Linux and either 
\texttt{LoadLibraryEx} or \texttt{CreateFileMappingA} and \texttt{MapViewOfFile} on Windows.

To map the shared library, the user only requires read-only access to the file containing the target page.
As the attacker process works on its own mapping of the shared library, all addresses are observed relative to the start of the shared library.
Hence, security mechanisms such as Address Space Layout Randomization (ASLR) have no effect on our attack.

To determine whether or not a page is in the page cache, we rely on the operating-system provides APIs to query the page cache. 
On Linux, this API is provided by the \texttt{mincore} system call.
\texttt{mincore} expects the base address and length of a memory area and returns a vector indicating for each page whether it is in the page cache or not. 
On Windows, there are two variants which are discussed as follows. 

\subsection{Windows Process Working-Set State}
On Windows, every process has a working set which is a very small subset of the page cache.
We cannot query the page cache directly as on Linux but instead we focus on the working set.
While this makes determining the cache state more complex, the following eviction is much easier and faster (\cf~\cref{sec:workingseteviction}).
On Windows, we rely on the \texttt{QueryWorkingSetEx} system call.
This function takes a process handle and an array specifying the virtual addresses of interest as arguments.
It returns a vector of structures which, if the page is part of the working set, contain various information about the corresponding pages.
In contrast to the official documentation~\cite{WinAPI}, the \texttt{QueryWorkingSetEx} system call only requires the \texttt{PROCESS\_QUERY\_LIMITED\_INFORMATION} permission. 
By default, the attacker process has this permission for handles of other processes of the same user and even for some processes with a higher integrity level (as part of the generic execute access)~\cite{MSDN2018}.
We devise two different variants to determine whether or not a page is in the working set of a process based on the return value of the \texttt{QueryWorkingSetEx} system call.

\parhead{Variant 1: Low Share Count and Attacker-Readable} 
The \texttt{ShareCount} represents the number of processes that have this page in their working set.
It is one of the members in the structure returned by \texttt{QueryWorkingSetEx}.
Unfortunately, the value is capped to 7 processes, \ie if more processes have the page in their working set, the number remains 7.
However, as the working-set size is limited to \SI{1.4}{\mega\byte} by default, this rarely happens for a page.
In fact, most pages in the page cache have a \texttt{ShareCount} of 0 due to the small working-set sizes.
With this variant, we do not need any permissions for other processes. 
Hence, we can mount the attack even across users without restrictions.

\parhead{Variant 2: High Share Count or Not Attacker-Readable} 
If the \texttt{ShareCount} is 7 or larger, we cannot gain any information by calling \texttt{QueryWorkingSetEx} on our own process.
Instead, we can use \texttt{QueryWorkingSetEx} directly on the victim process, \ie the attacking process must have the \texttt{PROCESS\_QUERY\_LIMITED\_INFORMATION} permission for the victim process handle.
As \texttt{QueryWorkingSetEx} takes virtual addresses, we need to figure out the virtual address.
This is not a problem if pages from shared files are targeted (\eg shared libraries) as they are typically mapped to the same virtual address in different processes.
However, if the pages are not shared, \ie not attacker-readable, \texttt{QueryWorkingSetEx} still leaks information if the virtual address is known.
Hence, we can use \texttt{QueryWorkingSetEx} to determine directly whether the target page is in the working set of the victim process.

\subsection{Spatial and Temporal Granularity} 
One limitation of our attack is the coarse spatial granularity of \SI{4}{\kilo\byte}, \ie one page. This is identical to a recent attack on TLB entries~\cite{Gras2018TLB} and similar to the DRAMA attack~\cite{Pessl2016,Wang2017leaky} on a single-channel DDR3 system, which has the same spatial granularity (one \SI{4}{\kilo\byte} page).
The spatial granularity of the DRAMA attack increases with the number of banks, ranks, channels, and processors.
It is \SI{2}{\kilo\byte} on dual-channel systems up to Haswell, and \SI{1}{\kilo\byte} on more recent dual-channel systems.
If a target region contains other frequently used data, the signal-to-noise ratio decreases in our attack just as it does for the DRAMA attack.
However, this just increases the number of measurements an attacker has to perform.

The temporal granularity of the DRAMA attack is constrained by the time it takes to run one or two rounds of \FlushReload, which is around \SI{300}{\nano\second}~\cite{Pessl2016,Wang2017leaky}.
The temporal granularity of our attack is constrained by the time the system call consumes, which we observed to be \SI{2.04}{\micro\second} on average for \texttt{mincore} with a standard error of \SI{20}{\nano\second}, and \SI{465.91}{\nano\second} on average for \texttt{QueryWorkingSetEx} with a standard error of \SI{0.20}{\nano\second}.
Hence, on Linux, it is only \SIx{6.8} times lower than the DRAMA attack, and on Windows only \SI{55}{\percent} lower than the DRAMA attack.
Thus, our attack can be used as a reasonable replacement for the hardware-dependent DRAMA attack.
However, as we describe in \cref{sec:eviction}, the eviction limits how often an attacker can measure, \ie $6.7$ times per second on Linux and $223$ times per second on Windows.

\subsection{Alternatives to \texttt{mincore} and \texttt{QueryWorkingSetEx}}
As an alternative to \texttt{mincore} on Linux, we also investigated whether it is possible to mount the same attack using \texttt{procfs} information, namely \texttt{/proc/self/pagemap}.
However, \texttt{/proc/self/pagemap} only shows the information from the page translation tables.
As operating systems commonly use lazy page mapping, the page is in practice not mapped into the attacker process and thus, the information in \texttt{/proc/self/pagemap} does not change.
Furthermore, as a response to Rowhammer attacks~\cite{Seaborn2015BH}, access to \texttt{/proc/self/pagemap} was first restricted and nowadays it is often not accessible by unprivileged processes.

As a more generic alternative to \texttt{mincore} and \texttt{QueryWorkingSetEx}, we investigated the timing of pagefaults as another source of information.
Accessing a page may trigger a pagefault.
Measuring the time it takes to handle the pagefault reveals whether it was a soft pagefault, mapping a page already present in the page cache, or a regular pagefault, loading data from the disk.
The timing differences we observed there are easy to distinguish, with 1 to 2 orders of magnitude between the two cases.
In our remote attack we exploit these timing differences.
However, this makes page cache eviction more difficult as the accessed page is now the least-recently used one.

Finally, as stated in \cref{sec:threatmodel}, our local attacks are entirely attack hardware-agnostic.
Hence, we cannot use any timing differences in our local attacks.

\section{Page Cache Eviction}\label{sec:eviction}\label{sec:eviction_sota}
In this section, we discuss how page cache eviction can be implemented efficiently on Linux and Windows systems.
Page cache eviction is the process of accessing enough pages in the right way such that a target page is evicted.
We show that we improve over state-of-the-art eviction algorithms by $1$ to $2$ orders of magnitude, enabling practical side-channel attacks through the page cache for the first time.

Less efficient variants of page cache eviction have been used in previous work~\cite{Holen2017,Gruss2018Rowhammer}.
Holen~\etal\cite{Holen2017}  generates a large amount of data, simply exhausting the physical memory.
Using this approach it takes \SI{8}{\second} or more to evict a target page on Linux.
Furthermore, when reproducing their results we observed severe stability issues, constantly leading to crashes and system lock-ups during eviction.
The technique presented by Gruss~\etal\cite{Gruss2018Rowhammer} takes \SI{2.68}{\second} on Linux to evict a target page.
On Windows, their technique is slower, with an average execution time of \SI{10.1}{\second}.
State-of-the-art microarchitectural side-channel attacks have a higher temporal resolution by more than \SIx{6} orders of magnitude~\cite{Yarom2014,Pessl2016,Maurice2017Hello}.
Hence, we can conclude that page cache eviction, as done in previous work, is far too slow for side-channel attacks with a relevant frequency. 
We solve this problem by combining the technique from \cref{sec:step1} with efficient page cache eviction on Linux (\cref{sec:ev_linux_efficient}) and process working-set eviction on Windows (\cref{sec:workingseteviction}).

\subsection{Efficient Page Cache Eviction on Linux}\label{sec:ev_linux_efficient}
The optimal cache eviction for the attacker would evict only the target page of the victim, without affecting other cached pages.
Hence, our idea is to mostly access pages which are already in the page cache (to keep them there) and also access a few non-cached pages in order to evict the target page.

In a feasibility analysis, we measured how many pages an attacker can locate inside the page cache. 
On our test system, we had \SI{1040542} files accessible to the attacker program, amounting to \SI{77}{\giga\byte} of disk space.
We found that less than \SI{1}{\percent} of the files had pages in the page cache, still amounting to \SI{68}{\percent} to \SI{72}{\percent} of the total page cache pages.
This information is all available to an unprivileged attacker using system calls like \texttt{mmap} and \texttt{mincore}.
The attacker creates a long list of all pages currently in the page cache.
The attacker also creates a list of further pages that could be loaded into the page cache to increase memory pressure.
Both lists can be updated occasionally to reflect changes in the system memory use.
The attacker adapts the amount of pages accessed in these two lists to achieve efficient cache eviction.

This is done by creating 3 eviction sets:

\parhead{Eviction Set 1}
These are pages already in the page cache, used by other processes. 
To keep them in the page cache, a thread continuously accesses these pages while also keeping the system load low by using \texttt{sched\_yield} and \texttt{sleep}.
Consequently, they are among the most recently accessed pages of the system and eviction of these pages becomes highly unlikely.

\parhead{Eviction Set 2}
These are pages not yet in the page cache.
Using \texttt{mincore}, we can check whether the target page was evicted, and stop the eviction immediately, reducing the eviction runtime.
Pages in this eviction set are randomly accessed, to avoid repeated accesses and thus any similarity to the pages in eviction set 1 for the replacement algorithm.

\parhead{Eviction Set 3}
If swapping is disabled, we use another eviction set, namely non-evictable pages, \eg dynamic content.
These pages are only created and filled with content, but never again read or written.
As they cannot be swapped, they block a certain amount of memory, reducing the required eviction-set size.
This reduces the runtime of the eviction significantly.
Still, this introduces no stability issues, as we always keep a large amount of pages ready for immediate eviction, \ie the previous 2 eviction sets.

\parhead{Alternative Approaches and Optimizations}
We investigated whether the file system influences the attack performance. 
For our tests, we used \texttt{ext4} as a file system.
We compared the attack performance by running our attack on \texttt{XFS} and \texttt{ReiserFS}.
However, we only found negligible timing differences.

We also investigated whether the use of the \texttt{madvise} and \texttt{posix\_fadvise} system calls on Linux can improve the attack performance. 
These system calls allow a programmer to provide usage hints for a given memory or file range to the kernel.
The advice \texttt{MADV\_DONTNEED} indicates that the process will not access the specified pages any time soon again, whereas the advice \texttt{MADV\_WILLNEED} indicates that the process will soon access the specified pages again.
Thus, the operating system will evict the corresponding pages from the page cache.
We found that marking the target page as \texttt{MADV\_DONTNEED} and all eviction set pages as \texttt{MADV\_WILLNEED} was often
ignored by the kernel, which ignores these hints unless the process exclusively owns the pages (\texttt{madvise}) or when
no other process has the file mapped (\texttt{posix\_fadvise}). 
Still, this allows to use \texttt{posix\_fadvise} on files regardless how frequently they are accessed, \eg via \texttt{read()}, as long as they are not mapped. 
Hence, we are able to mount a covert channel by using \texttt{posix\_fadvise} on a file which was not mapped by any (other) process, instead of eviction. 

\subsubsection{Evaluation}\label{sec:evicteval}
We measured the precision and recall of our eviction by monitoring a periodic event which was triggered every second.
The page cache eviction using all 3 eviction sets simultaneously achieves an average runtime of \SI{149}{\milli\second} ($\sigma=\SI{1.3}{\milli\second}$) on average and an F-Score of \SI{1.0}.

Hence, while the temporal resolution of our attack is generally \SI{2.04}{\micro\second} on Linux, the rate at which events can be observed in practice is lower.
The reason is that, if the event occurs, eviction is necessary, and thus, the temporal resolution for events with a higher frequency is limited to \SI{149}{\milli\second} on average.
This still allows capturing more than \SIx{6} keystrokes per second, enough to capture keystrokes accurately for most users~\cite{Schwarz2018KeyDrown}.
In this case, the temporal resolution of the DRAMA attack is 6 orders of magnitude higher~\cite{Pessl2016,Wang2017leaky}.

The temporal resolution is also significantly higher than that of page-deduplication attacks.
The frequency at which page deduplication happens is lower the more memory the system has, and has in use, and the less power the device should invest in deduplication.
In practice deduplication happens every 2 to 45 minutes, depending on the system configuration~\cite{Gruss2015dedup}.
Hence, our attack has an at least 800 times higher temporal resolution than the best page-deduplication attacks.

\parhead{Limitations}\label{sec:linux_limitations}
One obvious limitation of our approach is that the target page has to be in the page cache.
However, as detailed in \cref{sec:pagecache}, virtually all pages used by user programs end up in the page cache, even dynamically allocated ones.

On Linux, the page must also be accessible to the attacker, \eg file-backed memory such as binary pages, shared library pages, or other files.
This is exactly the same requirement (and limitation) of \FlushReload attacks~\cite{Yarom2014,Irazoqui2015Lucky,Gruss2015Template,Inci2016,Lipp2016,Irazoqui2016Cross}.
Other microarchitectural attacks, \eg \PrimeProbe, may not have this requirement but usually have other similarly constraining requirements, such as knowledge of the physical address which is difficult to obtain in practice~\cite{Maurice2017Hello}.
Page-deduplication attacks also do not have this limitation, but they face other limitations such as a significantly lower temporal resolution and, more recently, that page deduplication is mostly disabled or limited to deduplication within a security domain~\cite{WindowsServerDeduplication, RedHatKSM, Vanderveen2016}.
On Windows, we do not have this limitation, \ie we can also attack dynamically allocated memory on Windows.

Due to the nature of the exploited side channel, our attack comes with clear limitations.
Like other cache attacks, the side channel experiences noise if the target location is not only used by the event the attacker wants to spy on but also other events.
This is the same limitation as for any other cache side-channel attack~\cite{Yarom2014,Gruss2015Template}.

Another limitation which frequently poses a problem in hardware cache attacks is prefetching~\cite{Yarom2014,Gruss2015Template}.
Unsurprisingly, software again implements the same techniques as hardware.
When accessing the SSD, the Linux kernel reads ahead to increase the performance of file accesses.
If not specified otherwise, the readahead window is 32 pages large, \cf \texttt{/sys/block/sda/queue/read\_ahead\_kb}.
This is similar to the adjacent line prefetcher and the streaming prefetcher in hardware.
Whenever a cache miss occurs, the adjacent line prefetcher always fetches the sibling cache line region into the cache, \ie the adjacent \SI{64}{\byte}.
Whenever a second cache miss within a page occurs, the streaming prefetcher reads ahead of the cache miss and reads up to \SI{512}{\byte} (\ie 8 cache lines) into the cache.
Gruss~\etal\cite{Gruss2015Template} noted that this limits their attack to a small number of memory locations per page.
The same limitations apply to our work, \ie monitoring multiple pages within a 32-page window can be noisy.
However, we found that this still leaves a multitude of viable attack targets.
To avoid triggering the prefetcher, we add the pages surrounding the target page to the eviction set 1, \ie we reduce their chance of being evicted, in order to avoid all noise from prefetching, as no other page from this range will be accessed.

Finally, the attacker process can, of course, only perform measurements and evictions when it is scheduled.
Hence, scheduling can introduce false negatives into our attack.
Again, this is also the case for hardware cache attacks~\cite{Maurice2017Hello}.

Compared to previous work, we improve the state-of-the-art for page cache eviction by a factor of more than $16$ and additionally avoid cache eviction in most cases (\cf \cref{sec:step1}).
With these two building blocks, we are able to mount practical attacks as demonstrated in the following sections.
The ideal target for our attack is a function or data block which is used at frequencies below $8$ times per second, but where a temporal resolution of \SI{2}{\micro\second} can leak a sufficient amount of information.
Furthermore, our ideal target resides on a page which is mostly accessed for this function or data block, and not for unrelated functions or data.

\subsection{Process Working-Set Eviction on Windows}\label{sec:workingseteviction}
As previous page cache eviction techniques~\cite{Holen2017,Gruss2018Rowhammer} are too slow to mount generic side-channel attacks, we pursue a different approach on Windows.
Windows has per-process working sets~\cite{WinAPI}, which (by default) are constrained to a size between \SI{100}{\kilo\byte} and \SI{1.4}{\mega\byte}~\cite{WinAPI}.
Hence, we evict a page from the process working set rather than from the page cache.
Our results show that the runtime of the eviction is on par with eviction in hardware cache attacks.

We use process working-set eviction in both, covert channels and side-channel attacks. 
For a covert channel, the sender can add pages to the working set, \eg by accessing them.
To evict pages, we use an unintended behavior of \texttt{VirtualUnlock} that comes from a programming error~\cite{MSDN2018}.
Calling \texttt{VirtualUnlock} on a page which is not locked evicts it directly from the working set.
For reasons of backward-compatibility, the behavior was never changed~\cite{MSDN2018}.
Additionally, pages which are only read in one of the processes can be locked, so that they are never removed from the working set.
This way, arbitrary information can be encoded into the \texttt{ShareCount} of the page cache pages -- up to 3 bits exist, which allows 7 sharers.
Hence, we can transmit arbitrary information without any special privileges (as long as the receiver is not constrained by an App Container).
The default maximum working-set size is \SI{1.4}{\mega\byte}.
As the page size is \SI{4}{\kilo\byte}, that is, there are at most 345 page slots in the working set by default~\cite{WinAPI}.
Hence, we can exploit self-eviction (from the working set) for the side channel, which can happen frequently with a little heavy memory pressure because of the small working-set size.
Pages that are not accessed are evicted from the working set, but remain in RAM and mapped in the process.
However, we can speed up eviction by reducing the victim process' working-set size using \texttt{SetProcessWorkingSetSize} on the other process~\cite{WinAPI}.
The lowest possible value for the maximum working-set size is 13 pages (\SI{52}{\kilo\byte}).

\subsubsection{Evaluation}
We found that \texttt{VirtualUnlock} has a success rate of \SI{100}{\percent} over several million tests.
The average time to evict a page from the process working set with \texttt{VirtualUnlock} is \SI{4.48}{\milli\second} with a standard error of \SI{3.6}{\micro\second}.

Similarly to Linux (\cf \cref{sec:ev_linux_efficient}), the higher runtime of the eviction has a local influence on the temporal resolution of our attack.
Generally, the temporal resolution of our attack on Windows is \SI{466}{\nano\second}, which is only \SI{55}{\percent} lower than the temporal resolution of the DRAMA attack~\cite{Pessl2016,Wang2017leaky}.
The eviction on Windows via \texttt{VirtualUnlock} consumes \SI{4.48}{\milli\second}, limiting the temporal resolution for high-frequency events to \SI{4.48}{\milli\second} on average.
Thus, locally the temporal resolution of the DRAMA attack is 4 orders of magnitude higher than the temporal resolution of the DRAMA attack~\cite{Pessl2016,Wang2017leaky}. 
Again, this is fast enough for inter-keystroke timing attacks~\cite{Schwarz2018KeyDrown,Monaco2018}.

While prefetching posed a relevant limitation on Linux, it is no problem on Windows.
On Windows, features like SuperFetch fetch memory into the page cache, acting like an intelligent hardware prefetcher or speculative execution.
Indeed, SuperFetch speculatively prefetches pages from the disk into the main memory, based on similar past usage, \eg same time of day, same sequence of applications started~\cite{Schmid2007}.
However, these pages are not added to the working set of any process. 
Thus, our side channel remains entirely unaffected by these Windows features.
This makes the side channel very well suited for inter-keystroke timing attacks~\cite{Schwarz2018KeyDrown,Monaco2018}.

\parhead{Limitations}
Our attack on Windows has clear limitations, mainly introduced by the permissions required by the attacker.
More specifically, the \texttt{SetProcessWorkingSetSize} system call requires the \texttt{PROCESS\_SET\_QUOTA} permission on the process handle~\cite{WinAPI}.
By default, the attacker process has this permission for handles of other processes of the same user running on the same or a lower integrity level.
Processes with a higher integrity level, \eg processes running with Administrator privileges, cannot be attacked using this system call~\cite{MSDN2018}.
The \texttt{VirtualUnlock} only works on our own process and requires no permissions.
Also, noise is again a limitation, which exists for both the Linux and the Windows variant of our attack, but this is again also true for any other cache side-channel attack~\cite{Yarom2014,Gruss2015Template,Maurice2017Hello}.
In our tests, we were always able to reliably evict the page from the victim's working set indicating very low error rates.

\section{Local Attacks}\label{sec:local}
In this section we present and evaluate our local attacks.
The temporal resolution naturally scales with the performance of the system.
We perform all performance evaluations on recent systems with multiple gigabytes of RAM, with off-the-shelf mid-class consumer SSDs (\eg transfer rates above $\SI{250}{\mega\byte/\second}$~\cite{Tallis2018}).
For our tests on Linux, we have swapping disabled. 
This is recommended with recent processors (\eg Haswell or newer) and to reduce disk wear~\cite{DebianSSD,Horn2017Swap,Crowthers2011}.
Disabling swapping allows for a better comparison with related work which also focuses on such recent systems~\cite{Pessl2016,Wang2017leaky,Irazoqui2016Cross,Lipp2016}.

\subsection{Covert Channel}\label{sec:covert}
To systematically evaluate the page cache side channel, we adapt different state-of-the-art hardware cache attacks to it and demonstrate that they achieve a comparable performance.
In this section, we cover the first example, a covert channel between two processes additionally isolated by running them in different Firejail sandboxes~\cite{Firejail}.
The strongly isolated sender process sends a secret file from a restrained environment to a receiver process which can forward the data to the attacker.

As evicting a page is comparably slow (\cf \cref{sec:eviction}), and checking the state of a page is comparably fast (\cf \cref{sec:step1}), it is optimal to reduce the number of evictions.
Hence, it is more efficient to transmit multiple bits at once.
We took this into account for the design of our covert channel. 
We follow the basic principle of hardware cache covert channels~\cite{Maurice2017Hello,Pessl2016,Gruss2016Flush,Maurice2015C5}.
First, a large shared file (\eg a shared library) is mapped read-only into the address space of the sender and receiver process.
As described in \cref{sec:attack}, we use \texttt{mmap} for this purpose on Linux.
On Windows, we use \texttt{CreateFileMappingA} and \texttt{MapViewOfFile} for the same purpose.

The covert channel works by accessing or not accessing specific pages. 
We use two pages to transmit a `READY' signal and one page to transmit an `ACK' signal.
The remaining pages up to the end of the file are used as data transmission bits.
The two `READY' pages are used alternately to avoid any race conditions in the protocol between the transmission of two subsequent messages.
On Windows, we use two `READY' pages and two `ACK' pages, for the two transmission directions.

The present state of each page of the mapped file (\cf \cref{sec:step1}) corresponds to one bit of the message.
Hence, the size of the file defines the maximum message size of a single transmission.
To avoid the prefetcher, we only allow a single access in a region of 32 pages.
If the file has a size $S$, the (maximum) message size is computed as $w=\frac{S}{4096 \cdot 32} \textnormal{bits}$.
For instance, on Linux, Firefox' \texttt{libxul.so} or Chromium's \texttt{chromium-browser} binaries are more than \SI{100}{\mega\byte} large.
Similarly, large files can also be found on Windows.

These large files allow transmitting more than $3200$ bits in a single message including the 3 pages required for the control channels.
To avoid the introduction of noise, the attacker can skip noisy pages, \ie pages which are also accessed by other system activity.
By combining pages from multiple shared libraries, the attacker can easily find a significantly higher number of pages that can be used for transmissions, leading to very large message sizes $w$.
The pages are numbered from $0,1,..,i,..,w$, \ie it is not relevant which file they belong to.
Instead of a static list of files to check, the attacker could also use a dynamic approach and a jamming-agreement protocol~\cite{Maurice2017Hello}.

To exchange a message, the sender first checks the present state of the `ACK' page (\cf \cref{sec:step1}).
If the `ACK' page is present, the sender knows the receiver is ready for the next transmission.
The sender then evicts (\cf \cref{sec:eviction}) any pages that are mapped, \eg from previous transmissions.
After that, the sender reads the next $w$ bits ($w$ is the message size) from the secret to transmit.
If the $i$-th bit is set, page $i$ page is accessed.
Otherwise, page $i$ is not accessed.
As soon as the sender is done with accessing the data transmission pages, it accesses the currently to-be-set `READY' page, to signal the receiver to start reading the message.

On the other side, the receiver first waits until a `READY' page is present.
As soon as it is set, the receiver reads the message by analyzing the present state of the pages of the memory mapped files.
After that, the receiver accesses the `ACK' page again to inform the sender that it is ready for the next message.

While above protocol is implemented with \texttt{mmap}, \texttt{mincore} (\cf~\cref{sec:step1}), and page cache eviction (\cf~\cref{sec:ev_linux_efficient}) on Linux, we use a slightly different mechanism on Windows as we only work with working-set eviction (\cf~\cref{sec:workingseteviction}).
On Windows, we lock pages in the working set which should always remain in the working set, \ie the `READY' and `ACK' bit pages of the sender and the receiver process on the corresponding receiving side.
Additionally, we increase the minimal working-set size so that none of the pages we use are removed from the working set.
We temporarily add pages into the working set by accessing them and remove pages surgically from the working set by calling \texttt{VirtualUnlock}.
Hence, the covert channel information is perfectly (no information loss) stored in the page cache in the \texttt{ShareCount} for the shared pages.
Using \texttt{QueryWorkingSetEx} the receiving side can read the \texttt{ShareCount} and decode the information that was encoded in the page cache.

\parhead{Performance Evaluation}
We tested the implementation by transmitting random messages between two processes.
The test system was equipped with an Intel i5-5200U processor, \SI{8}{\giga\byte} DDR3-1600 RAM, and a \SI{256}{\giga\byte} Samsung SSD.

For the tests on Linux, we used Ubuntu 16.04 with kernel version 4.4.0-101-generic.
We observed transmission rates of up to \SI{9.69}{\kilo\byte/\second} with an average transmission rate of \SI{7.04}{\kilo\byte/\second} with a standard error of \SI{0.18}{\kilo\byte/\second}.
We did not observe any influence by the core or CPU scheduling, which is not surprising, as both the system calls and the page cache eviction can equally run on any core or CPU.
We observed a bit-error rate of less than \SI{0.00003}{\percent}.
We also evaluated the covert channel in a cross-sandbox scenario using Firejail~\cite{Firejail}.
Firejail was configured to prevent all outgoing inter-process communication, deny all network traffic, and only allow read access to the file system.
We did not observe any influence from running the covert channel in isolated Firejail sandboxes.
This is not specific to Firejail but works identically on other sandbox and container solutions that utilize the host system page cache, \eg Docker if configured accordingly.

For the tests on Windows, we used two different hardware setups with fully updated Windows 10 installations.
On the Intel i5-5200U system, we observed transmission rates of up to \SI{152.57}{\kilo\byte/\second} with an average transmission rate of \SI{100.11}{\kilo\byte/\second} with a standard error of \SI{0.79}{\kilo\byte/\second} and a bit-error rate below \SI{0.000006}{\percent}.
On a second system, an Intel i7-6700K with a SanDisk Ultra II 480GB SATA SSD (running Ubuntu 19.04 with a 4.18.0-11-generic kernel), we observed transmission rates of up to \SI{278.16}{\kilo\byte/\second} with an average transmission rate of \SI{273.44}{\kilo\byte/\second} with a standard error of \SI{0.23}{\kilo\byte/\second}, again with a bit-error rate below \SI{0.000006}{\percent}.

For a performance comparison in a similar cross-CPU scenario, Pessl~\etal\cite{Pessl2016} reported an error rate of \SI{0.4}{\percent} for their DRAMA covert channel, albeit with a channel capacity of \SI{74.5}{\kilo\byte/\second} which is much slower than our Windows-based covert channel, but faster than our Linux-based covert channel.
Wu~\etal\cite{Wu2014} presented a cross-CPU covert channel which achieves a channel capacity of \SI{93.25}{\byte/\second}.
Hence, our Linux covert channel outperforms this one by two orders of magnitude and our Windows covert channel even by three to four orders of magnitude.
In particular, the covert channel on the i7-6700K test system can even compete with \FlushReload and \FlushFlush covert channels which require specific hardware (Intel processors) and shared memory~\cite{Gruss2016Flush}.
Thus, we conclude that our covert channel can very well compete with state-of-the-art hardware-component-based covert channels.
Yet, our covert channel works regardless of the presence of these leaking hardware components.

\subsection{Authentication UI Redress Attack}\label{sec:redress}
In this section, we present a user-interface redress attack~\cite{Rydstedt2010,Niemietz2012,Chen2014,Bianchi2015,Ren2015,Fratantonio2017} which relies on our side channel as a trigger.
The basic idea is to detect when an interesting window is opened and to place an identically looking fake window over it.
This can be so stealthy that even advanced users do not notice it~\cite{Fratantonio2017}.
However, to achieve this, the latency between the original window opening and the fake window being placed over it must be very low.
Fortunately, our side channel provides us with exactly this capability, regardless of any other information leakage.
Note that the operating systems authentication windows may be protected.
However, other password prompts, \eg for password managers, browsers, and mail clients, are usually unprotected and can be targeted in our attack.

We use our side channel to detect when a root authentication window on Ubuntu 16.04 is displayed.
We detect this with a latency of \SI{2.04}{\micro\second} on average, and it does not take us longer to make our fake window visible and move it on top of the real window.
The user now types in the root password in our fake window.
Depending on the attacker capabilities, the attacker can either forward the password to the real window or simply close the fake window after the password was entered.
In the latter case, the user would see the original authentication window afterwards and likely think that the password was rejected on the first try, \eg because of a typing error occurred.

To identify binary pages which are used when spawning the root authentication window, we performed an automated template attack (\cf \cref{sec:keystroke}).
Note that the template attack is performed on an attacker-controlled system with identical software installed.
Hence, the attacker can take arbitrary means (\eg side-channel attacks or a debugger) to find interesting memory locations that can be exploited on the victim system.
The attacker first runs a debugger-based or cache-based template attack~\cite{Gruss2015Template} to identify binary regions that handle the corresponding event.
In a second run, the attacker templates with our page cache side-channel attack.
In our specific case, the result of the templating was that the strongest leakage is page 2 in the binary file \texttt{polkit-gnome-authentication-agent-1}.
Hence, on the victim system, the attacker simply uses the previously obtained templates to mount the attack.

Mounting the same attack on Windows 10 works even better.
Here, the latency is only \SI{465.91}{\nano\second}, which is clearly not perceivable for a human.
Also, unsurprisingly, we found that fake windows can be created on Windows just as on Linux.

Events like authentication windows and password prompts are very well suited for our attack due to the low frequency in which they occur.
This also makes the automated templating for leaking pages less noisy.

\subsection{Keystroke Timing Attack}\label{sec:keystroke}
In this section, we present an inter-keystroke-timing attack~\cite{Song2001,Ristenpart2009,Zhang2009,Gruss2015Template,Monaco2018} on keyboard input in the root authentication window on Ubuntu 18.04.
To mount a keystroke timing attack, we first identify pages that are loaded into the page cache when the user presses a key using a template attack~\cite{Gruss2015Template} (\cf \cref{sec:redress}).
We target the Ubuntu 18.04 authentication window, where the user types in the root password.
In the template attack, we identified page 14 of \texttt{libgksu2.so.0.0.2} as a viable target page.

\begin{figureA}[t]{keystrokes}[Values returned by the page cache side channel during a password entry on Linux (top) and while typing in an editor on Windows (bottom). On Windows we observe key up and key down events due to the page selected and the high attack frequency achievable. In both cases, there is no noise between the keystrokes.]
 \centering
\begin{subfigureA}[t]{\hsize}{keystroke_linux}
\centering
\begin{tikzpicture}
\begin{axis}[
mlineplot,
style={font=\footnotesize},
xlabel={Time [seconds]},
ylabel={Value},
width=1.0\hsize,
xmin=-0.25,
xmax=10.75,
ymax=1.5,
ymin=0,
ytick={0,1},
nodes near coords,
nodes near coords align={anchor=north,text height={8pt}},
point meta=explicit symbolic,
height=2.5cm
]
\addplot+[only marks,mark options={draw=black,fill=black}, mark=*] table[x=time,y=retval,meta=letter,col sep=comma] {keystrokes.csv};
\end{axis}
\end{tikzpicture}
\vspace{-0.5cm}
\end{subfigureA}

\begin{subfigureA}[t]{\hsize}{keystroke_windows}
\centering
\begin{tikzpicture}
\begin{axis}[
mlineplot,
style={font=\footnotesize},
xlabel={Time [seconds]},
ylabel={Value},
width=1.0\hsize,
xmin=-0.25,
xmax=9,
ymax=1.5,
ymin=0,
ytick={0,1},
nodes near coords,
nodes near coords align={anchor=north,text height={8pt}},
point meta=explicit symbolic,
height=2.5cm
]
\addplot+[only marks,mark options={draw=black,fill=black}, mark=*] table[x=time,y=retval,meta=letter,col sep=comma] {keystrokes_win.csv};
\end{axis}
\end{tikzpicture}
\end{subfigureA}
\end{figureA}

\cref{fig:keystrokes} shows two attack traces of a password entry, one on Linux (\cref{fig:keystroke_linux}) and one on Windows 10 (\cref{fig:keystroke_windows}) in \texttt{notepad.exe}.
We obtain identical traces on Windows when running the attack on Firefox.
Note that on Linux, for an extremely fast typing person, we could miss some keystrokes, \ie false negatives can occur.
However, we can gather these traces multiple times and combine the information from multiple traces to recover highly accurate inter-keystroke timings~\cite{Schwarz2018KeyDrown,Monaco2018}.
For Windows, the temporal resolution is much higher, far below the timing variations of a human~\cite{Schwarz2018KeyDrown,Monaco2018}, allowing us to reliably detect and report all inter-keystroke timings including key down and key up events.

When running the side-channel attack on an idle system for one hour, we did not observe a single false positive, neither on Windows nor on Linux.
This is not surprising, if the memory region is used by unrelated events we would have already seen such noise in the template phase.
However, as the attacker can and will choose the memory region based on the templating, the attacker chooses memory regions which are not really used by any unrelated events. 
Thus, in the optimal case, the selected memory region is completely noise-free.
In such a case, there is no functionality in the operating systems that could lead to false positives due to spurious cache hits.
Running the attacker binary inside a Firejail sandbox~\cite{Firejail} had no measurable influence on the accuracy of the attack.

\subsection{PHP Password Generation}\label{sec:token}
The PHP \texttt{microtime} function returns the current UNIX timestamp in microseconds.
It is carelessly used by some frameworks to initialize the PHP pseudo-random number generator (PRNG) before it is used in cryptographic operations or to generate temporary passwords~\cite{Zhang2014,Argyros2012,Esser2008}.
This is known as a bad practice and considered insecure, not least due to side-channel attacks~\cite{Zhang2014}.
During our research we found that the popular phpMyFAQ framework~\cite{Rinne2018} still relies on this approach.\footnote{We responsibly disclosed this vulnerability to the developers of phpMyFAQ who issued a patch following our recommendation.}

We mount our page cache attack on the main PHP binary (7.0.4-7ubuntu2), on the function \texttt{zif\_microtime}.
This function is read-only and shared with any unprivileged process including the attacker.
In our case, the function resides on page \texttt{0x1b9} (\SIx{441}) of the binary.
By monitoring this page, we can determine the return value of \texttt{microtime} at the initialization of the PRNG.
Based on this, we can reconstruct any password generated based on the same PRNG initialization, as the password generation algorithm is also publicly available. 

Due to the large variance on the runtime of PHP scripts, we only detected an access to the \texttt{microtime} function with an accuracy of $\pm \SI{1.5}{\milli\second}$.
However, this is practical to brute force the range of remaining possible return values. 
On a newer PHP version (7.0.30-0ubuntu0.16.04.1), we observed an average difference of $\pm \SI{2.0}{\milli\second}$.
Thus, we have to try around \SIx{4000} different passwords in the real-world attack.
We confirmed that in \SI{85}{\percent} of the test runs, the real password of the user was among the \SIx{4000} generated passwords from the attacker.
Hence, also in this scenario, our page cache side channel can compete with state-of-the-art attacks~\cite{Zhang2014}.

Our attack also works on Windows.
However, as the main source of noise is the varying runtime of PHP, the accuracy is not measurably better on Windows.

\subsection{Oracle Attacks}\label{sec:oracle}
Our side channel also allows implementing padding- or length-oracle attacks.
For instance, a password or token comparison using \texttt{strcmp} forms a length oracle.
If the attacker can place the string on the page boundary, the attacker can measure at which byte of the string the comparison terminated.
By manipulating the string, the attacker can figure our the correct password or token.

We verified that this attack is practical in a small proof-of-concept program.
The attacker passes the string through an API to the victim process.
By using our page-cache- or working-set-based side channel we can determine whether the second page was loaded into the page cache or added to the working set.
If this was the case, the attacker learns that the bytes on the first page were guessed correctly.

As the attacker can fully control the frequency of the measurements here and can repeat the attack, we observed no cases where we could not successfully leak the secret.

\section{Remote Attack}\label{sec:remote}
For our remote attack we have to distinguish soft pagefaults, \ie just mapping the page from the page cache, and regular pagefaults, \ie page cache misses, over a network connection.
In this scenario, two physically separated processes wish to communicate with each other. 
The sender process runs on a server and has access to information that the attacker wants to have.
However, it is unprivileged, firewalled, and possibly sandboxed, so it cannot reach any network resources or expose files for remote access. 
However, the server exposes multiple files to the public internet, \eg over a web server.
We also assume that the sender process has read permissions to these files, \eg Apache has world-readable permissions on files in the web server root directory by default. 
The receiver process runs on a remote server, measuring the remote access latency to pages in these public files.
Hence, the sender process can encode the information in the page cache state of these pages. 

\begin{figureA}[t]{cache_hit_miss_dist}[Timing histogram of the remote covert channel with a \SI{100}{\kilo\byte} file (25 pages).]
\centering
\begin{tikzpicture}
\begin{axis}[
style={font=\footnotesize},
xlabel={Latency [ms]},
ylabel={Frequency},
height=3.5cm,
scaled x ticks=false,
width=\hsize,
xmin=5,
xmax=15,
]
\addplot+[no marks,thick] table[x=time,y=hits,col sep=comma] {remote_timing.csv};
\addplot+[no marks,densely dotted,thick] table[x=time,y=misses,col sep=comma] {remote_timing.csv};
\legend{Hits,Misses}
\end{axis}
\end{tikzpicture}
\end{figureA}

\parhead{Page Cache Hits and Misses}
Of course, a remote attacker cannot invoke \texttt{mincore} to check which pages are in cache, so the attacker needs to rely on timing. 
Hence, we first try to distinguish cache hits and misses over the network, similarly to the related work in~\cite{Tiwari2018,Schwarz2018netspectre}, by performing remote accesses with and without clearing the page cache. 
We also ensured that there was no other intermediary network caching or proxy caching active by passing appropriate HTTP headers to the server.
\Cref{fig:cache_hit_miss_dist} shows the frequencies of remote access latencies for various cached and uncached accesses; the figure shows that cache hits
can be distinguished from cache misses. Here, the mean access time was \SI{8.4}{\milli\second} for cache hits and \SI{14.2}{\milli\second} for cache misses to access a file with 25 pages (around \SI{100}{\kilo\byte}).
The latency differences between cache hits and misses grow with the number of pages accessed.
Hence, we use larger files for the subsequent remote attacks.

\subsection{Covert Channel Protocol}
\begin{figure}
\centering
\includegraphics[width=\hsize]{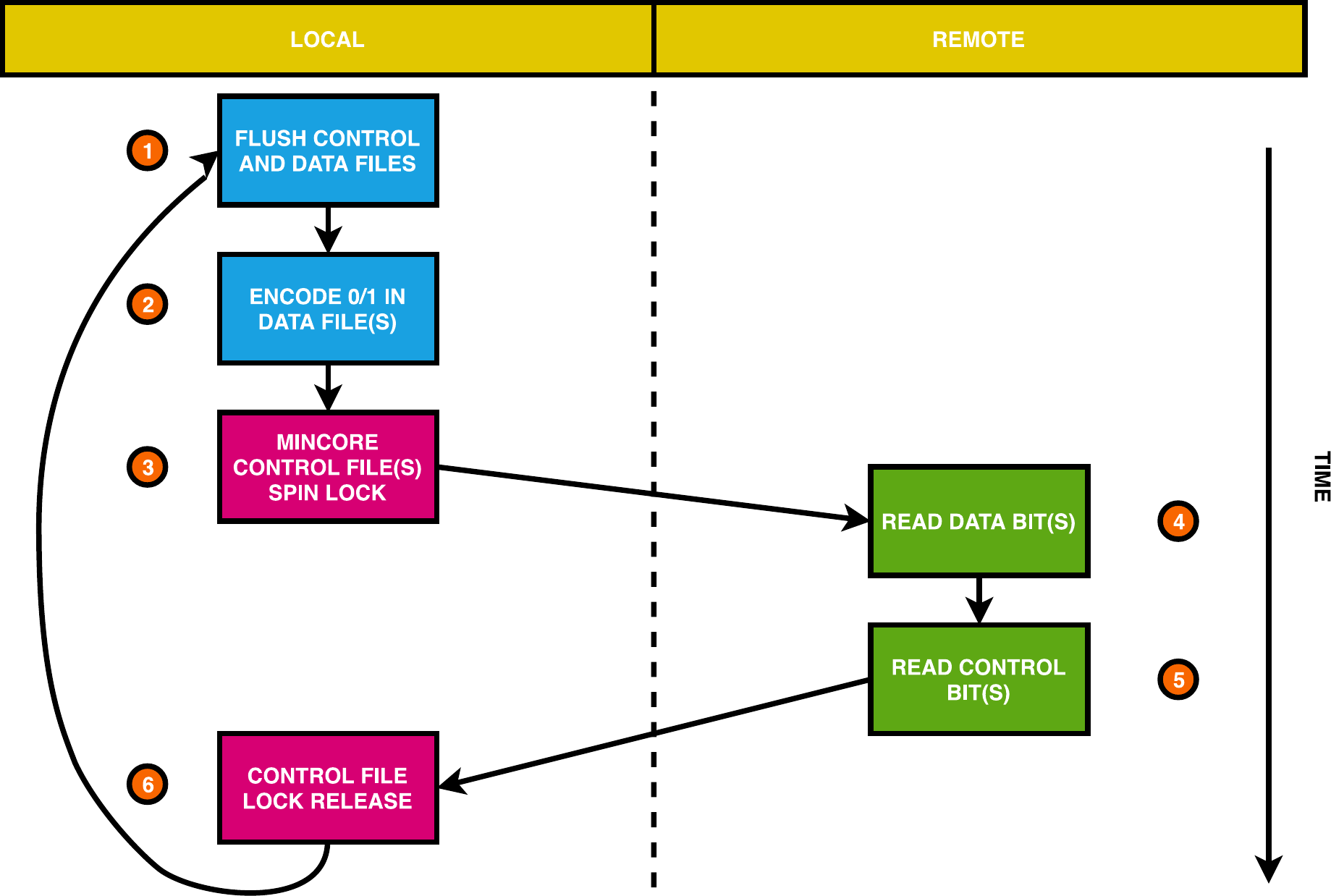}
\caption{Illustration of the web server covert channel.}
\label{fig:remote_covert_channel} 
\end{figure}

\Cref{fig:remote_covert_channel} depicts how the two processes communicate over the covert channel. 
The local sender process is an unprivileged (possibly sandboxed) malware that encodes secret data from the victim machine into page cache hits and misses, and the remote receiver process decoding the secret data after measuring the remote access latency. 
For this, the sender process uses one file to encode data, and another file for synchronization (control file). 
The sender process first evicts both the data and control files from the file system cache (Step 1) using \texttt{posix\_fadvise} on a rarely used file, \ie a file which is not currently locked in memory by another process.
Note that the attacker could also use any other means of page cache eviction as described in~\cref{sec:eviction}.
It then encodes one bit of information in the data file (Step 2) by either bringing it into the page cache by reading the file (encoding a `1'), or not bringing it into the cache (encoding a `0'). 
After encoding, the sender waits for the control file to be read by the remote process (Step 3).
For this, the sender uses \texttt{mincore} on the control file in a loop, checking how many of the file's pages are in the page cache.
In our case, the sender waits until \SI{80}{\percent} of the file are cached, indicating that the remote attacker accessed it.

\begin{figureA}[t]{remote_covert_channel_res}[Transmitting a sequence of alternating `0's and `1's by accessing a \SI{10}{\mega\byte} file (2560 pages). A threshold can distinguish the two cases.]
\centering
\begin{tikzpicture}
\begin{axis}[
style={font=\footnotesize},
ylabel={Latency [ms]},
xlabel={Sequence Number},
height=3.5cm,
width=\hsize,
xmin=2000,
xmax=2049,
ytick={0.095,0.1,0.105,0.11,0.115},
yticklabels={95,100,105,110,115},
]
\addplot+[] table[x=packet,y=time,col sep=comma] {covert.csv};
\addplot[thick, dashed, red] coordinates {(2000,0.105)(2099,0.105)};
\end{axis}
\end{tikzpicture}
\end{figureA}

The receiver process measures the access latency, inferring the bits the sender process was trying to transmit (Step 4). 
In our experiments, the access time threshold that demarcated a `0' from a `1' was set to \SI{105}{\milli\second} for our hard drive experiments, as illustrated in \Cref{fig:remote_covert_channel_res}.

Immediately after the receiver process accessed the data file, it also accesses the control file (Step 5), to let the sender know the next bit can be transmitted now.
The sender then continues at Step 1 again.
This happens until the sender has transmitted all bits of secret information.

\parhead{Evaluation}
Our experimental setup involved two separate, but geographically close, machines, \ie a network distance of 4 hops.
The victim machine was running the Linux Mint (kernel version 4.10.0-38) on an AMD A10-6700 with \SI{8}{\giga\byte} RAM and a \SI{977}{\giga\byte} hard drive. 
The victim machine exposed two files to the network, \texttt{data.jpg} (\SI{10}{\mega\byte}) and \texttt{control.jpg}, used as the data and control files respectively. 
The remote machine was also running Linux Mint (kernel version 4.13.0-37) on an Intel Core i7-7700 with \SI{16}{\giga\byte} RAM and a \SI{219}{\giga\byte} SSD.

For the evaluation, we transmitted \SI{4000} bits from the local machine to the remote machine multiple times. 
The transmission took \SI{517}{\second} on average, which corresponds to an average bit rate of \SI{7.74}{\bit/\second} and an average bit error rate of \SI{0.2}{\percent}. 
This is a higher bit rate than several other remote covert channels~\cite{Crosby2009,Cock2014timing,Schwarz2018netspectre}. 
The bit rate can be further increased by encoding information through more than one file, which is realistic given the vast number of files most web servers today have. 
To increase stealthiness, the attacker may choose to access the two files from different IPs, as the sender process is agnostic to this. 

As our covert channel relies on timing differences, we also repeated our experiments on a machine with an SSD.
Distinguishing a page cache hit from a page cache miss through timing over the network, could be more difficult as the timing differences can be smaller. 
To overcome this, we simply use a larger image file (\SI{30}{\mega\byte}, 7680 pages) to amplify the timing difference. However, this meant that the load latency threshold that demarcates a read hit and miss would need to be scaled up similarly from the previous experiment, and was set to \SI{300}{\milli\second} for the experiments on SSDs.
Furthermore, we reduce the geographical distance between attacker and victim to 2 network hops. 
The victim server was running on a machine with Linux Mint on an Intel Core i7-7700 (kernel version 4.13.0-37) with \SI{16}{\giga\byte} RAM and a recent off-the-shelf \SI{219}{\giga\byte} SSD, and the attacker machine was the same as before. 
The transmission of \SI{4000} bits, now takes \SI{1298}{\second} on average, giving us an average bit rate of \SI{3.08}{\bit/\second} at an average bit error rate of \SI{0.35}{\percent}.
Hence, this remote timing covert channel is also possible on a machine with an SSD. 

Our proof-of-concept implementation could be further optimized to yield a higher transmission rate, to mount the attack over a greater geographical distance, or to use smaller files, simply by repeating measurements for each single bit~\cite{Schwarz2018netspectre}.
In our proof-of-concept we did not repeat any measurements to obtain a single bit, again indicating the high capacity of this remote covert channel.

\subsection{Remote Side Channel}
Similarly to our local side-channel attacks, we could also mount remote side-channel attacks exploiting the page cache.
This information could be used to determine whether certain pages or scripts have been recently accessed~\cite{Tiwari2018}.
However, in practice it is difficult to evict the cache remotely and eviction can be tricky without the information from the local system.
Furthermore, controlling the working set via a huge number of remote file accesses will make the attack very conspicuous,
though it may still be practically effective for opportunity-based attacks (\eg password reset pages) such as those presented in Section~\ref{sec:token}.

\section{Countermeasures}\label{sec:countermeasures} 
Our side-channel attack targets the operating system page cache via operating system interfaces and behavior.
Hence, it clearly can be mitigated by modifying the operating system implementation.

\parhead{Privileged Access}
The \texttt{QueryWorkingSetEx} and \texttt{mincore} system calls are the core of our side-channel attack.
Requiring a higher privilege level for these system calls stops our attack.
The downside of restricting access to these system calls is that existing programs which currently make use of these system calls might break.
Hence, we analyzed how frequently \texttt{mincore} is called by any of the software running on a typical Linux installation.
We used the Linux \texttt{perf} tools to measure over a 5 hour period whenever the \texttt{sys\_enter\_mincore} system call is called by any application.\footnote{We use \texttt{sudo perf stat -e 'syscalls:sys\_enter\_mincore' -a sleep 18000} for this purpose.}
During these 5 hours a user performed regular operations on the system, \ie running various work-related tools like Libre Office, gcc, Clion, Thunderbird, Firefox, Nautilus, and Evince, but also non-work-related tools like Spotify.
The system was also running regular background tasks during this time frame.
Surprisingly, the \texttt{sys\_enter\_mincore} system call was not called a single time.
This indicates that making the \texttt{mincore} system call privileged is feasible and would mitigate our attack at a very low implementation cost.

On Windows, there are multiple possible solutions to mitigate our attacks by adapting the privileges required for the system calls we use.
First of all, it is questionable why a process can obtain working-set information of another process via \texttt{QueryWorkingSetEx}.
Especially, as this contradicts the official documentation~\cite{WinAPI}.
Second, the share count information could be omitted from the struct returned by \texttt{QueryWorkingSetEx} as it exposes information about other processes to the attacker.
The combination of these two changes mitigates all our attack variants on Windows.

We responsibly disclosed our findings to Microsoft, and they acknowledged the problem and will roll out these changes in Windows 10 19H1.
Specifically, Windows will require \texttt{PROCESS\_QUERY\_INFORMATION} for \texttt{QueryWorkingSetEx} instead of \texttt{PROCESS\_QUERY\_LIMITED\_INFORMATION} to prevent lesser privileged processes from directly obtaining working set information.
Microsoft also follows our second recommendation of omitting the share count information, to prevent indirect observations on working set changes in other processes.

It was also surprising that Windows allows changing the working-set size for another process.
If this would be restricted, it would be much more difficult to reliably evict across processes.
The performance of our covert channel would decrease if \texttt{VirtualUnlock} did not have the ``feature'' that it removes pages from the working set if they are not locked.

Alternative approaches like page locking, signal burying, or disabling page sharing are likely not practical for most use cases or impose significant overheads.

\parhead{Preventing Efficient Eviction while Increasing the System Performance}
On Windows, we used working set eviction instead of page cache eviction as on Linux.
We verified that the approach we used on Linux, \ie page cache eviction, also works on Windows.
However, it performs much worse than on Linux and optimizing the eviction appeared to be far more tricky.
One reason for this is that with working-set-based algorithms, processes cannot directly influence the eviction probability for pages owned by or shared with other processes~\cite{Denning1968,Denning1980,Carr1981}.
On Linux, we are only able to evict pages efficiently because we can trick the page replacement algorithm into believing our target page would be the best choice for eviction.
The reason for this lies in the fact that Linux uses a global page replacement algorithm, \ie an algorithm which does not distinguish between different processes.
Global page replacement algorithms have been known for decades to allow one process to perform a denial-of-service on other processes~\cite{Denning1968,Denning1980,Carr1981,Russinovich2012}.

Working-set algorithms deplete these denial-of-service situations and they also increase the general system performance by making more clever choices for eviction candidates~\cite{Denning1968,Denning1980,Carr1981}.
Hence, switching to working-set algorithms on Linux, as on Windows~\cite{Russinovich2012}, makes our attack less practical.
We can also transfer this insight to hardware caches: If hardware caches would use replacement algorithms that guarantee fairness in a similar way, attacks like \PrimeProbe would not be possible anymore, because the attacker would rather evict its own cache lines, rather than the one required by the victim process.
This is a larger change, but it might make remote attacks that rely on page cache eviction less practical.

\section{Conclusion}\label{sec:conclusion} 
We have demonstrated a variety of local and remote attacks against the page cache used in modern operating systems, thereby highlighting a new source for side- and covert channels that is hardware and timing agnostic.
On the local front, we have demonstrated a high-speed cross-sandbox covert channel, a UI redressing attack triggered by a side channel, a keystroke-timing side channel, and password-recovery side channel from a vulnerable PHP script.
On the remote front, we have shown that forgoing hardware agnosticism permits a low profile covert channel from a local malicious sender, and a higher profile side channel.
The severity of this attack surface is exacerbated by the variety of isolation techniques that share the page cache, including regular Unix processes, sandboxes, Function-as-a-Service platforms, managed language runtimes, web browsers, and even select remote processes.
Stronger permissioning, as we recommend, will help against some of our local attacks.

\section*{Acknowledgments}
We want to thank James Forshaw for helpful discussions on COM use cases and Simon Gunacker for early explorative work on this topic.
Daniel Gruss and Michael Schwarz were supported by a generous gift from ARM and also by a generous gift from Intel.
Ari Trachtenberg and Trishita Tiwari were supported, in part, by the National Science Foundation under Grant No. CCF-1563753 and Boston University’s Distinguished Summer Research Fellowship, Undergraduate Research Opportunities Program, and the department of Electrical and Computer Engineering.
Any opinions, findings, and conclusions or recommendations expressed in this paper are those of the authors and do not necessarily reflect the views of the funding parties.

{\footnotesize \bibliographystyle{acm-url}
  \bibliography{main,additional}}

\end{document}